# Forecasting in the Fog: Real-Time versus Revised-Data Evidence on Machine Learning's Edge over the Phillips Curve

**Louis Agyekum**[1,*], **Obed Obese**[2]

[1] Department of Economics, University of Ottawa, Ottawa, ON, K1N 6N5, Canada

[2] Department of Economics, Carleton University, Ottawa, ON, Canada

* Correspondence: agyekumlouis9@gmail.com

## Abstract

ML forecasts of inflation are almost universally trained on fully revised data, even though real-time forecasters never have such data, and reported feature importances are typically computed in-sample, conflating predictive relevance with retrospective fit. This paper tests whether the ML advantage over the Phillips curve documented in Agyekum (2026) survives when models are trained and evaluated on real-time (ALFRED) vintages rather than revised series, and examines whether SHAP feature-importance rankings are an artifact of in-sample estimation. Using 2000-2026 U.S. data on unemployment, CPI and PCE inflation, payrolls, real GDP, and the 10-year-2-year Treasury spread, vintage-consistent panels are built for four traditional models (random walk, AR(1), Phillips curve, ADL-OLS) and four ML models (Random Forest, Gradient Boosting, Elastic Net, SVR), re-estimated recursively at 3-, 6-, and 12-month horizons (208, 206, 204 forecasts). Real-time/revised accuracy differences are small and, apart from one exception at 6 months (Gradient Boosting vs. Phillips curve, DM = -1.671, p = 0.097), indistinguishable under Diebold-Mariano tests; Gradient Boosting alone shows consistent positive skill at longer horizons. The random walk remains a strong short-horizon benchmark, consistent with the puzzle in Agyekum et al. (2026) for exchange rates. Using walk-forward, out-of-sample SHAP, a Random Forest on revised data assigns dominant importance to PCE inflation (mean |SHAP| = 0.778, rank 1 of 9), while on real-time data it assigns PCE negligible importance (0.039, rank 6), relying instead on current CPI (0.834 vs. 0.223). This twenty-fold swing, larger than the in-sample estimate, is invisible to point-forecast metrics and shows the model's PCE reliance is substantially a hindsight artifact. An RSI summarizes the accuracy gap by model and horizon, with implications for auditing ML inflation forecasts.

# 1. Introduction

In the first quarter of 2022, the U.S. Bureau of Economic Analysis reported real GDP growth of −1.4 percent. One month later, the estimate was revised to −1.5 percent; by the third and final estimate, it stood at −1.6 percent (BEA, 2022). A forecaster relying on any of these three release dates was working with a different figure for the same economic quarter and would not learn, for months, the value that economic historians would eventually treat as definitive for that period. This is not an isolated feature of one volatile quarter: nearly every headline U.S. macroeconomic series, including GDP, payrolls, and, to a lesser degree, inflation, is revised repeatedly after its first release, sometimes for years, and occasionally by amounts sufficient to alter the qualitative characterization of economic conditions at a given point in time.

This creates a significant, underappreciated problem for the rapidly growing literature applying machine learning (ML) to macroeconomic forecasting. With few exceptions, published comparisons of ML against traditional econometric benchmarks, including studies by Agyekum (2026) on the Phillips curve and Nortey et al. (2025) on deep learning for inflation forecasting, are trained and evaluated on the final, fully revised vintage of each series. A forecaster operating in real time never has this advantage: at the moment a forecast is generated, only the data published up to that date are available. If part of ML's apparent forecasting advantage over simpler benchmarks derives from patterns in data that did not actually exist at the time a forecast was made, a subtle form of look-ahead bias is embedded in the standard evaluation design. That advantage is at least partly illusory, and any policy or investment decision predicated on it rests on an evaluation that could never be replicated by an actual real-time forecaster.

This paper addresses a direct question: does machine learning's forecasting advantage over the traditional Phillips curve survive when both model classes are trained and evaluated strictly on real-time, vintage-consistent data? The analysis draws on the ALFRED (ArchivaL Federal Reserve Economic Data) database, which records every historical revision to U.S. macroeconomic series and the date each revision became public, to reconstruct precisely what a forecaster would have known at each point in history. A fully recursive, re-estimated-every-month forecasting exercise is then implemented, a computationally demanding but methodologically necessary design that much of the existing literature, including several of the studies cited above, does not employ. It compares four traditional models against four ML models at three forecast horizons, under both real-time and revised data conditions, and formally assesses the statistical significance of every comparison using the Diebold–Mariano procedure.

The contribution of this paper is fourfold. First, it provides one of the earliest fully vintage-consistent, multi-model, multi-horizon comparisons of ML and traditional inflation forecasting for the post-pandemic U.S. economy, directly extending the SHAP-based Phillips curve comparison in Agyekum (2026) to a real-time setting with a substantially larger model zoo. Second, it formalizes a diagnostic measure, the Revision Sensitivity Index (RSI), defined as the gap between a model's real-time and revised-data forecast accuracy, and demonstrates, using Diebold–Mariano significance testing across thirty-six model-horizon comparisons, that this gap is generally small and statistically indistinguishable from zero, a finding informative about the external validity of the standard (revised-data) evaluation convention. Third, it shows that point-forecast accuracy conceals an important divergence in model behavior: SHAP interpretability analysis reveals that a revised-data-trained Random Forest relies heavily on PCE inflation as a predictor, whereas the identical model trained on real-time data makes little use of it, a hindsight artifact invisible to RMSFE-based comparisons. Fourth, the paper provides, as supplementary material, a fully reproducible open-source pipeline for constructing real-time vintage panels from the ALFRED API, intended to reduce the fixed cost of conducting this style of analysis for future researchers.

The remainder of the paper proceeds as follows. Section 2 reviews the real-time data literature and situates the present contribution within related recent work on ML macroeconomic forecasting. Section 3 describes the data and documents the extent of revision in each series. Section 4 formalizes the real-time panel construction, the eight forecasting models, the recursive evaluation design, and the statistical and interpretability testing procedures, each stated with explicit mathematical notation. Section 5 presents results. Section 6 discusses the findings and their implications for practice. Section 7 addresses limitations and robustness, and Section 8 concludes.

# 2. Related Literature

## 2.1 Real-Time Data and the Evaluation of Forecasting Models

The recognition that macroeconomic data revisions can materially affect both estimated models and forecast evaluations dates at least to Croushore and Stark (2001), who introduced the real-time dataset for macroeconomists that underlies the modern ALFRED database and demonstrated that econometric results can differ meaningfully depending on the data vintage used for estimation. Clements and Hendry (2005) formalize the general principle underlying the evaluation design adopted in the present study: a forecasting exercise is genuinely "real time" only if both the estimation sample and the model itself are re-derived at each forecast origin using the information available at that date, as opposed to the more common "pseudo-real-time" practice of using a fixed, most-recently-available data vintage throughout. Ellingsen, Larsen, and Thorsrud (2022) implement this distinction empirically in a nowcasting context, comparing real-time GDP nowcasts constructed from ALFRED vintages with news-based predictors, and find that the choice of data vintage can materially affect which predictors appear informative, a pattern documented independently via SHAP in Section 5.5 below.

A long-standing literature shows that simple, low-information forecasts are difficult for structural models to outperform. Atkeson and Ohanian (2001) showed that a naive random-walk forecast of next period's inflation, equal to this period's inflation, performed at least as well as several Phillips-curve specifications over the 1984–2000 period, a result often cited as marking the beginning of serious doubt about the Phillips curve's forecasting usefulness. Stock and Watson (2007) extended this line of inquiry by documenting that U.S. inflation became substantially harder to forecast, across model classes, after the mid-1980s Great Moderation, attributing the deterioration to a decline in the persistent, forecastable component of inflation relative to its transitory, unforecastable component. The results reported in Section 5.1, in which the random walk remains competitive with or superior to seven more sophisticated alternatives at short and medium horizons under both data conditions, are consistent with both studies and extend them into a real-time, machine-learning-augmented setting.

## 2.2 Machine Learning in Macroeconomic Forecasting

A more recent and rapidly expanding literature applies machine learning methods to macroeconomic forecasting problems traditionally dominated by linear time-series and structural models. Coulombe,

Leroux, Stevanovic, and Surprenant (2022) provide a systematic comparison showing that tree-based ensembles and regularized or neural methods can outperform classical linear approaches in nonlinear, high-dimensional settings, with gains concentrated at longer horizons and near business-cycle turning points. They caution that the size of these gains is sensitive to data construction, sample period, and model calibration choices, a caution directly relevant to our finding that ML's advantage is horizon-dependent and, at the 3-month horizon, statistically indistinguishable from the traditional Phillips curve.

Several recent studies inform the present analysis directly. Nortey et al. (2025) compare deep learning, machine learning, and traditional statistical models for inflation time-series forecasting and find that the relative advantage of more flexible model classes is both horizon- and regime-dependent, a finding corroborated in the present study: Gradient Boosting is the standout model at the 6- and 12-month horizons but offers no significant advantage at 3 months. Agyekum (2026) applies SHAP interpretability to a Phillips-curve-style comparison of machine learning and traditional time-series models for Canadian unemployment and inflation, finding that lagged inflation dominates unemployment as a predictor, except during the pandemic tail. The SHAP results reported in Section 5.5 extend this interpretability approach to the real-time/revised distinction, specifically, which that earlier study, in common with the rest of the literature, did not address.

Agyekum et al. (2026) document a closely related phenomenon in a different asset class: in expanding-window forecasts of the CAD/USD exchange rate, a naive random-walk benchmark remains extremely difficult for machine learning models to beat. SHAP analysis shows that short lags and recent rolling means dominate model predictions, consistent with near-random-walk behavior in the underlying series. The consistency between the exchange-rate results reported in that study and the inflation-forecasting results reported here, with the random walk winning or nearly winning in both applications, suggests that the phenomenon reflects a general property of highly persistent, near-unit-root macro-financial series rather than an artifact specific to either asset class, a point revisited in Section 6.

## 2.3 This Paper's Contribution Relative to the Literature

No existing study combines (i) a fully recursive, re-estimated-every-period real-time evaluation design constructed from genuine ALFRED vintage data, (ii) a model zoo spanning both traditional econometric and modern machine learning specifications, (iii) formal Diebold–Mariano significance testing of the real-time/revised accuracy gap itself, and (iv) SHAP-based interpretability comparison across data vintages, within a single, unified inflation-forecasting exercise. The closest antecedents to the present study each provide one or two of these four elements: Ellingsen et al. (2022) provide (i) for GDP nowcasting without (ii)–(iv); Coulombe et al. (2022) provide (ii) and partial (iii) using revised data throughout; and Agyekum (2026) and Agyekum et al. (2026) provide (ii) and (iv) using revised data and, for the exchange-rate study, data that does not undergo the kind of ex-post revision central to this paper's research question. The present study is, to the authors' knowledge, the first to bring all four elements together for U.S. inflation forecasting specifically.

## 3. Data and Real-Time Vintage Construction

### 3.1 Series and Sources

We use six monthly and quarterly U.S. macroeconomic series obtained from the Federal Reserve Bank of St. Louis's FRED and ALFRED databases, spanning January 2000 through mid-2026: the civilian unemployment rate (UNRATE), the Consumer Price Index for All Urban Consumers (CPIAUCSL), all-employee nonfarm payrolls (PAYEMS), real Gross Domestic Product (GDPC1), the Personal Consumption Expenditures price index (PCEPI, the Federal Reserve's preferred inflation gauge), and the 10-year-minus-2-year Treasury constant-maturity yield spread (T10Y2Y). For the first five series, we retrieve the complete ALFRED vintage history- every value ever published for every historical observation date, tagged with the date each revision became public. The Treasury spread is market-observed and is not subject to ex-post revision, so it is retrieved as a single-vintage series and treated as such throughout.

### 3.2 Revision Statistics

Table 1 documents how often each series is revised. Unemployment is revised relatively rarely (an average of 1.9 times per observation, with 58 percent of observations revised at all, mostly reflecting annual seasonal-factor updates). CPI inflation is revised more often (5.0 times on average) but, as Table 2 shows, by very small amounts. Nonfarm payrolls and real GDP, by contrast, are revised extensively: payroll observations are revised an average of 12.9 times, and GDP 10.3 times. In our sample, GDP revisions occasionally reflect large benchmark or chain-weighting updates rather than routine re-estimation; the largest single revision to year-over-year GDP growth was 15.1 percentage points, associated with a 2023 benchmark update to the national accounts. PCE inflation, though less commonly examined in the real-time literature than GDP or payrolls, is revised nearly as often as GDP (10.7 times on average) and by economically large amounts, a fact that becomes directly relevant to the SHAP results in Section 5.5.

*Table 1. Revision frequency by series, 2000–2026.*

| Series | Avg. revisions / obs. | Max. revisions | % ever revised |
|---|---|---|---|
| UNRATE (unemployment rate) | 1.87 | 5 | 58.2% |
| CPIAUCSL (CPI) | 4.98 | 7 | 97.2% |
| PAYEMS (nonfarm payrolls) | 12.92 | 22 | 99.7% |
| GDPC1 (real GDP) | 10.29 | 15 | 99.1% |
| PCEPI (PCE price index) | 10.68 | 18 | 99.7% |
| T10Y2Y (Treasury spread) | 1.00 | 1 | 0.0% |

Table 2 reports the magnitude of these revisions in year-over-year growth terms, comparing each series' first-released value with its most recently available ("final") value. CPI revisions are economically negligible (mean 0.001 percentage points, standard deviation 0.10pp), which is reassuring given that CPI inflation is both our forecast target and a key predictor. Payroll and PCE revisions are an order of magnitude larger (standard deviations of 0.40pp and 3.18pp, respectively), and GDP revisions are larger still, with a standard deviation of 4.51 percentage points, providing direct evidence that GDP and PCE are the series for which the real-time/revised distinction is most consequential.

*Table 2. Revision magnitude, year-over-year growth terms (final minus first release).*

| Series | Mean revision (pp) | Std. dev. (pp) | Max. abs. revision (pp) |
|---|---|---|---|
| CPIAUCSL | 0.001 | 0.095 | 0.354 |
| PAYEMS | −0.034 | 0.401 | 1.025 |
| GDPC1 | −1.762 | 4.513 | 15.130 |
| PCEPI | 1.436 | 3.181 | 11.148 |

**Figure 1.** *Levels and growth rates of the six macroeconomic series used in this study, final/revised vintage.*

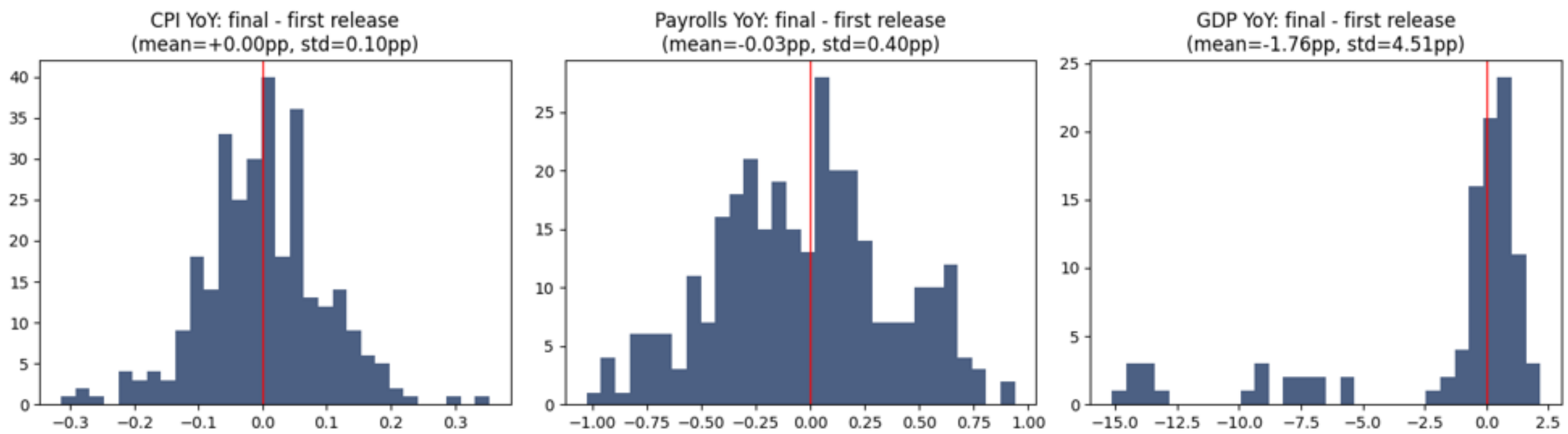

**Figure 2.** *Distribution of revision magnitudes for CPI, payrolls, and GDP (year-over-year growth terms).*

# 4. Methodology

This section formalizes the construction of the real-time and revised information sets, states each of the eight forecasting models in explicit mathematical notation, and details the recursive evaluation, significance testing, and interpretability procedures. Figure 3 summarizes the full analytical pipeline; each stage is described formally in the subsections that follow.

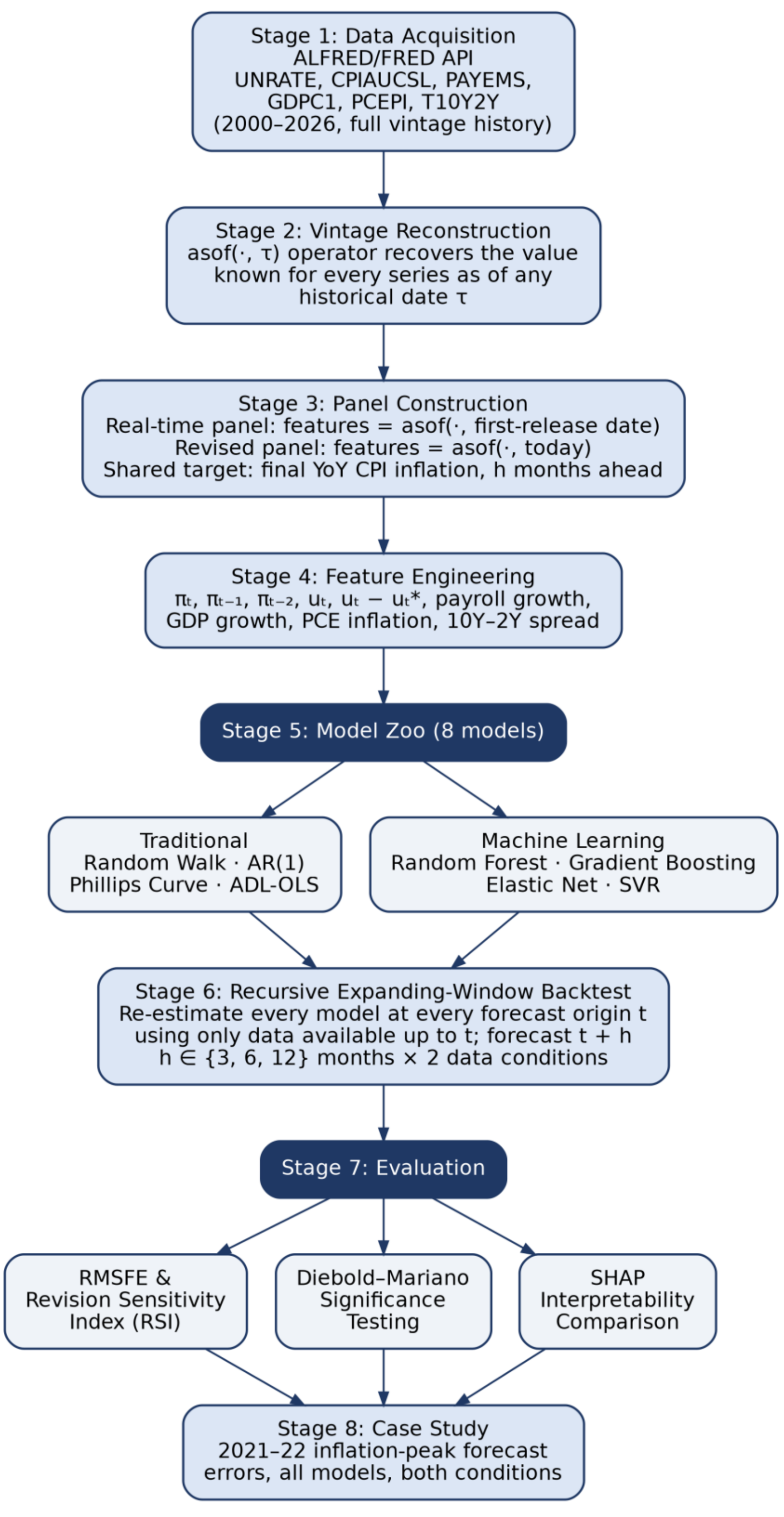

*Figure 3. The eight-stage analytical pipeline: from ALFRED vintage acquisition through recursive backtesting to the interpretability case study.*

## 4.1 Notation and the Real-Time Information Set

Let $y_t^{(v)}$ denote the value of a macroeconomic series for observation period $t$ as published in vintage $v$, and let $\mathrm{RTS}(v)$ denote the real-time-start date of that vintage, the date on which value $v$ became the current published figure. For any historical date $\tau$, define the as-of operator:

$$y_t^{asof(\tau)} = y_t^{(v^*)}, \qquad v^* = \mathrm{argmax}\{\mathrm{RTS}(v) \leq \tau : v \in \mathcal{V}(t)\} \tag{1}$$

which recovers the value of series $y$ that was actually known to a forecaster standing at date $\tau$, for every observation period $t \leq \tau$. Applying this operator with $\tau$ equal to the first-release date of each observation recovers the real-time information set; applying it with $\tau$ equal to the present recovers the fully revised ("final") series used throughout the wider literature.

All growth-rate transformations are computed in year-over-year (YoY) terms to avoid the base-year and level discontinuities associated with periodic benchmark revisions (Section 3.2). For a price or activity index $P_t$:

$$\pi_t = 100 \times \left(\frac{P_t}{P_{t-12}} - 1\right) \tag{2}$$

For each historical reference month $r$, we construct two parallel feature vectors. The real-time feature vector uses $asof(\cdot, \tau)$ evaluated at the date CPI inflation for month $r$ was first published; the revised feature vector uses $asof(\cdot, \tau)$ evaluated at the present. Both share an identical, fully revised target: the ex-post "true" realized inflation $h$ months later:

$$\mathbf{X}_t^{RT} = \mathbf{X}_t^{asof\left(\tau_t^{first}\right)}, \qquad \mathbf{X}_t^{REV} = \mathbf{X}_t^{asof\left(\tau^{today}\right)}, \qquad y_{t+h} = \pi_{t+h}^{final} \tag{3}$$

For each reference month we construct nine predictors: current CPI inflation and two lags $(\pi_t, \pi_{t-1}, \pi_{t-2})$, the unemployment rate $u_t$ and its gap from a trailing five-year (60-month) mean $u_t^*$ (a simple, time-varying natural-rate proxy computed using only data available within each panel's own information set), nonfarm payroll growth $g_t$, real GDP growth $y_t$, PCE inflation $p_t$, and the 10-year–2-year Treasury spread $s_t$.

## 4.2 Traditional Econometric Models

### 4.2.1 Random Walk

The random walk forecast assumes future inflation equals current inflation, with no re-estimation required:

$$\hat{\pi}_{t+h|t} = \pi_t \tag{4}$$

Despite its simplicity, this specification serves as the benchmark against which the real-time forecasting literature has repeatedly found more elaborate models struggle to add value (Atkeson & Ohanian, 2001).

### 4.2.2 Autoregressive Model, AR(1)

The AR(1) model estimates a single autoregressive coefficient on current inflation via ordinary least squares, re-estimated at every forecast origin using only training data available up to that point:

$$\hat{\pi}_{t+h|t} = \hat{\alpha} + \hat{\beta}\,\pi_t \tag{5}$$

### 4.2.3 The Accelerationist Phillips Curve

The traditional Phillips-curve specification regresses future inflation on current inflation and the unemployment gap:

$$\hat{\pi}_{t+h|t} = \hat{\alpha} + \hat{\beta}_1\pi_t + \hat{\beta}_2(u_t - u_t^*), \qquad u_t^* = \frac{1}{60}\sum_{i=1}^{60} u_{t-i} \tag{6}$$

This is the classic backward-looking, accelerationist formulation examined in Atkeson and Ohanian (2001) and in Agyekum (2026), and serves as the paper's primary traditional-model benchmark for the Diebold–Mariano comparisons in Section 5.3.

### 4.2.4 Autoregressive Distributed-Lag Model (ADL-OLS)

To give the traditional model class access to the same information set as the machine learning models, we also estimate a multivariate linear regression on the full nine-variable feature vector:

$$\hat{\pi}_{t+h|t} = \hat{\alpha} + \hat{\boldsymbol{\beta}}'\mathbf{X}_t, \qquad \mathbf{X}_t = [\pi_t, \pi_{t-1}, \pi_{t-2}, u_t, u_t - u_t^*, g_t, y_t, p_t, s_t]' \tag{7}$$

This specification remains linear econometrics estimated by ordinary least squares with no regularization, but it allows us to isolate whether any real-time/revised sensitivity observed in the machine learning models is attributable to the estimator's flexibility itself or simply to the inclusion of heavily revised predictors such as GDP and PCE growth.

## 4.3 Machine Learning Models

### 4.3.1 Random Forest

Random Forest (Breiman, 2001) is an ensemble of $B$ regression trees, each trained on a bootstrap resample of the training data with a random subset of features considered at each split, with predictions averaged across trees:

$$\hat{f}_{RF}(\mathbf{X}_t) = \frac{1}{B}\sum_{b=1}^{B} T_b\,(\mathbf{X}_t) \tag{8}$$

We use $B = 200$ trees, maximum depth 4, and a minimum of 5 observations per leaf, chosen to limit overfitting given the modest sample sizes available in each recursive training window.

### 4.3.2 Gradient Boosting

Gradient Boosting (Friedman, 2001) builds an additive ensemble sequentially, at each stage $m$ fitting a shallow regression tree $h_m$ to the negative gradient of the loss function with respect to the current ensemble prediction, and adding it to the ensemble with a shrinkage (learning rate) parameter $\nu$:

$$F_m(\mathbf{X}_t) = F_{m-1}(\mathbf{X}_t) + \nu\, h_m(\mathbf{X}_t), \qquad h_m \approx \text{argmin}_h \sum_i \left[-\frac{\partial L(y_i, F_{m-1})}{\partial F_{m-1}} - h(\mathbf{X}_i)\right]^2 \tag{9}$$

We use 150 boosting stages, maximum tree depth 2, and $\nu = 0.05$. Gradient Boosting is, as Section 5.1 shows, the only model in our comparison that consistently and substantially outperforms every other specification, including the random walk, at the 6- and 12-month horizons.

### 4.3.3 Elastic Net

Elastic Net (Zou & Hastie, 2005) is a regularized linear regression combining L1 (lasso) and L2 (ridge) penalties, providing feature selection alongside coefficient shrinkage:

$$\widehat{\boldsymbol{\beta}}_{EN} = \text{argmin}_{\boldsymbol{\beta}} \left\{\frac{1}{2n} \| \mathbf{y} - \mathbf{X}\boldsymbol{\beta} \|_2^2 + \lambda \left[\alpha \| \boldsymbol{\beta} \|_1 + \frac{1-\alpha}{2} \| \boldsymbol{\beta} \|_2^2\right]\right\} \tag{10}$$

We set the mixing parameter $\alpha = 0.5$ and regularization strength $\lambda = 0.1$, with all features standardized to zero mean and unit variance prior to estimation, as required for the penalty terms to be scale-comparable across predictors with very different natural units.

### 4.3.4 Support Vector Regression (SVR)

SVR (Drucker, Burges, Kaufman, Smola, & Vapnik, 1997) fits a function within an $\varepsilon$-insensitive tube around the observed targets, penalizing only residuals that exceed $\varepsilon$, and is estimated here with a radial basis function (RBF) kernel to capture non-linear relationships:

$$\min_{\mathbf{w},b} \frac{1}{2} \| \mathbf{w} \|^2 + C \sum_i (\xi_i + \xi_i^*), \qquad K(\mathbf{x}_i, \mathbf{x}_j) = \exp\left(-\gamma \| \mathbf{x}_i - \mathbf{x}_j \|^2\right) \tag{11}$$

We use $C = 1.0$, $\varepsilon = 0.1$, and the RBF kernel with scikit-learn's default bandwidth. SVR is, across every horizon and data condition examined, the weakest-performing model in our comparison (Section 5.1); we discuss this and its implications for hyperparameter sensitivity in Section 7.

## 4.4 Recursive Expanding-Window Evaluation Design

Both data conditions (real-time and revised) and all eight models are evaluated using a fully recursive, expanding-window design: at each monthly forecast origin t, every model is re-estimated from scratch using only the training observations available up to that point, then used to generate a single out-of-sample forecast for the corresponding target date t + h. The window then expands by one month, and the process repeats through the end of the sample. This is markedly more computationally intensive than the single-split or fixed-window designs common in the ML macroeconomic forecasting literature, but it is the methodologically correct choice for a genuine real-time exercise, following the general protocol of Clements and Hendry (2005): a forecaster's model itself, not merely its inputs, would have been re-estimated as new data arrived. We require a minimum of 60 months of training data before generating the first recursive forecast and a minimum of 84 months of underlying vintage history before a reference month is eligible for inclusion in either panel, yielding 208, 206, and 204 out-of-sample forecasts at the 3-, 6-, and 12-month horizons, respectively, spanning February 2007 through the end of the sample.

## 4.5 Forecast Accuracy Metrics

Forecast accuracy for model $m$ is summarized by the root-mean-squared forecast error over the $N$ recursive out-of-sample forecasts:

$$\text{RMSFE}_m = \sqrt{\frac{1}{N}\sum_{t=1}^{N}\left(\pi_{t+h} - \hat{\pi}_{t+h|t}^{(m)}\right)^2} \tag{12}$$

To directly quantify the sensitivity of each model's accuracy to the real-time/revised distinction, we define the Revision Sensitivity Index:

$$\text{RSI}_m = \text{RMSFE}_m^{RT} - \text{RMSFE}_m^{REV} \tag{13}$$

A positive RSI indicates that a model performs worse under real-time conditions than it would appear to based on the revised-data evaluation convention used throughout most of the literature, that is, that some of its apparent accuracy is attributable to hindsight. A negative RSI indicates the reverse.

Two further diagnostics are reported to characterize forecast accuracy in more detail than a single full-sample RMSFE permits. First, rolling RMSE is computed over a trailing 24-month window to visualize how relative forecast accuracy evolves over time, rather than collapsing the entire evaluation period into one number. Second, out-of-sample $R^2$ is computed relative to the random-walk benchmark, following the convention standard in the forecasting literature (e.g., Campbell & Thompson, 2008; Clark & West, 2007):

$$R_{OOS,m}^2 = 1 - \frac{\text{MSE}_m}{\text{MSE}_{RW}} \tag{14}$$

where $\text{MSE}_m$ and $\text{MSE}_{RW}$ denote the out-of-sample mean squared error of model $m$ and of the random-walk benchmark, respectively, over the identical recursive evaluation window. A positive $R_{OOS,m}^2$ indicates that model $m$ outperforms the random walk out of sample; a negative value indicates underperformance. Unlike in-sample $R^2$, this measure cannot be inflated by overfitting, since every forecast entering $\text{MSE}_m$ is generated from a model that had not observed the corresponding target at the time of estimation.

## 4.6 Diebold–Mariano Significance Testing

RMSFE and RSI values alone cannot establish whether an observed difference reflects a genuine, structural distinction between models or data conditions, as opposed to sampling variation in a finite evaluation sample. We therefore apply the Diebold and Mariano (1995) test to squared-error loss differentials. For two competing forecasts with errors $e_{1,t}$ and $e_{2,t}$, define the loss differential:

$$d_t = L(e_{1,t}) - L(e_{2,t}), \qquad L(e_t) = e_t^2 \tag{15}$$

The test statistic is the sample mean loss differential, standardized by its long-run variance, computed with a Newey–West-type correction at lag $h-1$ to account for the serial correlation mechanically induced by overlapping $h$-step-ahead forecast errors:

$$DM = \frac{\bar{d}}{\sqrt{\widehat{\text{Var}}(\bar{d})/N}}, \qquad \widehat{\text{Var}}(\bar{d}) = \hat{\gamma}_0 + 2\sum_{k=1}^{h-1}\hat{\gamma}_k \tag{16}$$

Because our sample sizes (204–208 observations) are modest relative to the asymptotic assumptions underlying the original Diebold–Mariano statistic, we apply the small-sample correction of Harvey, Leybourne, and Newbold (1997):

$$DM^{HLN} = DM \times \sqrt{\frac{N+1-2h+h(h-1)/N}{N}} \tag{17}$$

We run two families of comparisons at each horizon: each machine learning model against the Phillips curve under real-time conditions (testing whether ML's edge over the traditional benchmark is statistically genuine), and each model's real-time errors against its own revised-data errors (the direct test of whether the RSI is statistically distinguishable from zero). $DM^{HLN}$ is compared against the Student's $t$ distribution with $N-1$ degrees of freedom to obtain a two-sided p-value.

## 4.7 SHAP Interpretability

To assess whether a model's reliance on particular predictors is itself an artifact of the data vintage used for training, a question point-forecast accuracy metrics cannot address SHAP (SHapley Additive exPlanations; Lundberg & Lee, 2017) feature importances are computed for the Random Forest model at the 3-month horizon, separately for the real-time and revised data conditions. SHAP values are grounded in the cooperative game theory concept of the Shapley value (Shapley, 1953): for a prediction function $f$ and feature set $\mathcal{F}$, the contribution of feature $i$ is the value it adds averaged over all possible orderings in which features could be revealed to the model:

$$\phi_i(f) = \sum_{S\subseteq\mathcal{F}\setminus\{i\}} \frac{|S|!\,(|\mathcal{F}|-|S|-1)!}{|\mathcal{F}|!}[f(S\cup\{i\}) - f(S)] \tag{18}$$

where $f(S)$ denotes the model's expected prediction using only the subset of features $S$.

A critical methodological choice concerns which fitted model, f, is used to explain the start equation. The great majority of applications of SHAP to macroeconomic forecasting, including preliminary versions of the present analysis, fit a single model on the full estimation sample and compute SHAP values over that same sample: an in-sample interpretability exercise. This is potentially misleading for the same reason that in-sample goodness-of-fit is a poor guide to forecasting performance: a feature that appears important because the model has learned to exploit sample-specific noise need not be genuinely informative for forecasting new observations. To avoid this, feature importance in the present study is instead computed out-of-sample, using the same recursive expanding-window structure as the forecast evaluation itself. At each forecast origin start equation t in the recursive backtest, having fit model f sub t on training data available only up to start equation t (Section 4.4), the Shapley contribution of each feature is computed for that period's single out-of-sample test point bold cap X sub t using f sub t, the model that genuinely existed at that point in time, rather than a model fit on the full sample including data unavailable at start equation t. Out-of-sample feature importance for feature start equation i is then the sample mean of these per-period contributions across all start equation cap N recursive forecasts:ng data unavailable at $t$. Out-of-sample feature importance for feature $i$ is then the sample mean of these per-period contributions across all $N$ recursive forecasts:

$$\bar{\phi}_i^{OOS} = \frac{1}{N}\sum_{t=1}^{N}|\phi_i(f_t; \mathbf{X}_t)| \tag{19}$$

This procedure is computationally more demanding than the in-sample alternative, since it requires re-fitting and re-explaining a separate model at every recursive step rather than once, but it ensures that reported feature importances reflect genuinely out-of-sample interpretability rather than retrospective fit. Section 5.5 reports both the out-of-sample importances from equation (19) and, for comparison, the corresponding in-sample importances, to make the magnitude of the discrepancy between the two approaches directly visible. Feature rankings are compared across the real-time and revised data conditions under the out-of-sample procedure to detect hindsight-driven shifts in model behavior.

# 5. Results

## 5.1 Point-Forecast Accuracy

Tables 3–5 report RMSFE for all eight models across both data conditions at each forecast horizon, along with the Revision Sensitivity Index. Two patterns stand out. First, the random walk is a formidable benchmark at short and medium horizons: it achieves the lowest or near-lowest RMSFE among all eight models at h = 3 and h = 6 under both data conditions, consistent with the random-walk puzzle documented for exchange-rate forecasting in Agyekum et al. (2026) and with the classical finding of Atkeson and Ohanian (2001) that naive forecasts are difficult for Phillips-curve-style models to beat. Second, Gradient Boosting is the standout machine learning model at longer horizons, achieving the lowest RMSFE at h = 6 (0.907 real-time, 0.936 revised) and h = 12 (1.304 real-time, 1.451 revised). This is the only case in our results where an ML model clearly and consistently outperforms every traditional benchmark, including the random walk, a pattern consistent with the horizon-dependent ML advantage documented in Nortey et al. (2025) and Coulombe et al. (2022).

*Table 3. RMSFE by model and data condition, 3-month horizon (n = 208 evaluation months, Feb. 2007–Mar. 2026).*

| Model | Type | RMSFE (real-time) | RMSFE (revised) | RSI |
|---|---|---|---|---|
| Random walk | Traditional | 0.8093 | 0.8051 | +0.0042 |
| AR(1) | Traditional | 0.8314 | 0.8287 | +0.0027 |

| Model | Type | RMSFE (real-time) | RMSFE (revised) | RSI |
|---|---|---|---|---|
| Phillips curve | Traditional | 0.8391 | 0.8387 | +0.0004 |
| ADL-OLS | Traditional | 0.9706 | 0.9000 | +0.0706 |
| Random Forest | ML | 0.9169 | 0.8956 | +0.0213 |
| Gradient Boosting | ML | 0.8713 | 0.8552 | +0.0161 |
| Elastic Net | ML | 0.9054 | 0.8467 | +0.0588 |
| SVR | ML | 1.1230 | 1.0506 | +0.0724 |

*Table 4. RMSFE by model and data condition, 6-month horizon (n = 206 evaluation months, Feb. 2007–Aug. 2025).*

| Model | Type | RMSFE (real-time) | RMSFE (revised) | RSI |
|---|---|---|---|---|
| Random walk | Traditional | 1.2920 | 1.2895 | +0.0025 |
| AR(1) | Traditional | 1.3772 | 1.3777 | −0.0005 |
| Phillips curve | Traditional | 1.3995 | 1.4071 | −0.0075 |
| ADL-OLS | Traditional | 1.7312 | 1.4290 | +0.3022 |
| Random Forest | ML | 1.2752 | 1.3121 | −0.0368 |
| Gradient Boosting | ML | 0.9071 | 0.9363 | −0.0292 |
| Elastic Net | ML | 1.4901 | 1.3863 | +0.1039 |
| SVR | ML | 1.4224 | 1.3450 | +0.0774 |

*Table 5. RMSFE by model and data condition, 12-month horizon (n = 204 evaluation months, Feb. 2007–Jun. 2025).*

| Model | Type | RMSFE (real-time) | RMSFE (revised) | RSI |
|---|---|---|---|---|
| Random walk | Traditional | 2.1414 | 2.1439 | −0.0026 |
| AR(1) | Traditional | 2.0144 | 2.0146 | −0.0002 |
| Phillips curve | Traditional | 1.9355 | 1.9378 | −0.0023 |
| ADL-OLS | Traditional | 2.1702 | 1.7757 | +0.3945 |
| Random Forest | ML | 1.9061 | 1.8011 | +0.1050 |
| Gradient Boosting | ML | 1.3042 | 1.4514 | −0.1473 |
| Elastic Net | ML | 1.9956 | 1.7708 | +0.2248 |
| SVR | ML | 1.6775 | 1.6669 | +0.0107 |

## 5.2 Rolling RMSE and Out-of-Sample R²

A single full-sample RMSFE, as reported in Tables 3–5, can obscure substantial variation in relative forecast accuracy over time, and cannot by itself indicate whether a model's accuracy gain over the random-walk benchmark is large enough to be economically meaningful rather than a marginal statistical improvement. Figure 4 plots rolling 24-month RMSE for four representative models at h = 3 under the real-time condition. Two patterns are apparent. First, all four models exhibit a pronounced, common spike in rolling RMSE beginning in 2021, coinciding with the onset of the post-pandemic inflation surge; forecast difficulty during this episode dominates the full-sample RMSFE figures reported earlier and is not evenly distributed across the sample. Second, Random Forest tracks the random-walk benchmark closely for most of the sample but is *not* uniformly better or worse; the two lines cross repeatedly, underscoring why the aggregate RMSFE comparison in Table 3 shows no statistically significant difference between the two.

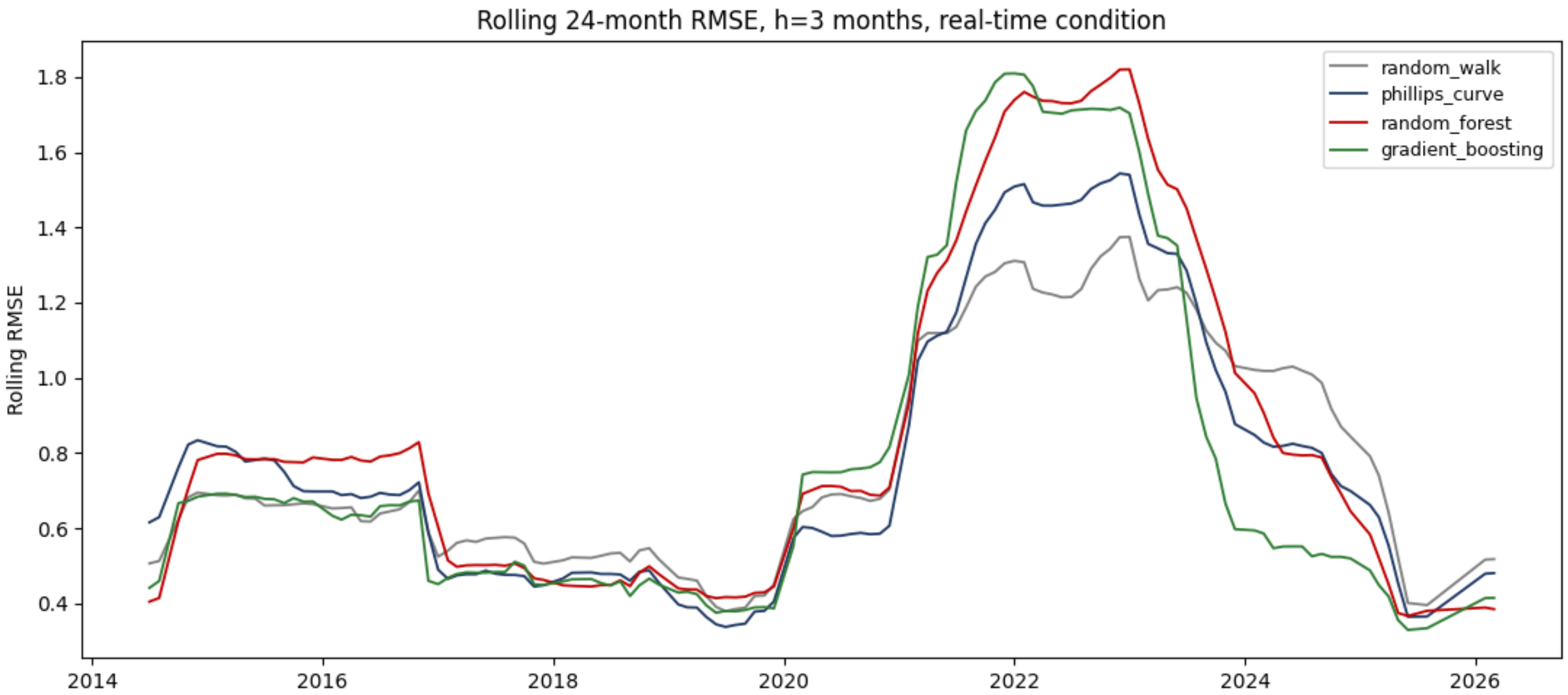

**Figure 4.** *Rolling 24-month out-of-sample RMSE by model, h = 3, real-time condition.*

Table 6 reports out-of-sample $R^2$ relative to the random-walk benchmark (equation 14) for every model, horizon, and data condition. The pattern corroborates and sharpens the RMSFE results in Section 5.1: at h = 3, every model, traditional and ML alike, has *negative* out-of-sample $R^2$, meaning none improves on the random walk out of sample at the shortest horizon, with SVR performing worst (−0.93 real-time). At h = 6, Gradient Boosting is the only model with a substantial positive out-of-sample $R^2$ (0.51 real-time, 0.47 revised), confirming it as the paper's sole example of genuine, out-of-sample-validated forecasting skill beyond the random walk. At h = 12, several models achieve positive out-of-sample $R^2$, with Gradient Boosting again the strongest (0.63 real-time, 0.54 revised), consistent with the general finding in the forecasting literature that ML methods add the most value at longer horizons (Coulombe et al., 2022).

*Table 6. Out-of-sample R² relative to the random-walk benchmark, by model, horizon, and data condition.*

| Model | Type | h=3 (RT) | h=3 (Rev) | h=6 (RT) | h=6 (Rev) | h=12 (RT) | h=12 (Rev) |
|---|---|---|---|---|---|---|---|
| Random walk | Traditional | 0.0000 | 0.0000 | 0.0000 | 0.0000 | 0.0000 | 0.0000 |
| AR(1) | Traditional | −0.0554 | −0.0596 | −0.1362 | −0.1415 | 0.1151 | 0.1170 |
| Phillips curve | Traditional | −0.0750 | −0.0852 | −0.1734 | −0.1907 | 0.1831 | 0.1831 |
| ADL-OLS | Traditional | −0.4384 | −0.2497 | −0.7954 | −0.2282 | −0.0272 | 0.3140 |
| Random Forest | ML | −0.2837 | −0.2377 | 0.0258 | −0.0354 | 0.2076 | 0.2942 |
| Gradient Boosting | ML | −0.1591 | −0.1284 | 0.5071 | 0.4728 | 0.6291 | 0.5417 |
| Elastic Net | ML | −0.2518 | −0.1061 | −0.3302 | −0.1558 | 0.1315 | 0.3178 |
| SVR | ML | −0.9256 | −0.7031 | −0.2120 | −0.0880 | 0.3863 | 0.3955 |

## 5.3 The Revision Sensitivity Index

The RSI is generally small in absolute magnitude for the purely autoregressive traditional models (random walk, AR(1), Phillips curve), rarely exceeding ±0.01 in either direction across all three horizons, unsurprising, since these models depend only on lagged inflation and the unemployment gap, both of which revise relatively little (Table 1). The RSI is larger and more variable for the multivariate models (ADL-OLS and all four ML models), which draw on payrolls, GDP, and PCE, the three most heavily revised series in our sample. ADL-OLS shows the largest and most consistent real-time penalty, with RSI rising from +0.071 at h = 3 to +0.395 at h = 12, indicating that an unregularized linear model using all nine features is the specification most damaged by the loss of hindsight a result consistent with the theoretical expectation that unregularized estimators are more prone to overfitting noisy, heavily-revised predictors than either shrinkage-based (Elastic Net) or ensemble-based (Random Forest, Gradient Boosting) alternatives. Among ML models, Elastic Net shows a similar, if smaller, pattern (RSI of +0.059 to +0.225 across horizons), while Gradient Boosting is the only model whose RSI is negative (i.e., real-time outperforming revised) at both h = 6 and h = 12, suggesting its regularized, shallow-tree structure may be comparatively robust to or even benefit from the noisier, less-revised real-time inputs.

## 5.4 Statistical Significance

Table 7 reports Diebold–Mariano test results for the twelve machine-learning-versus-Phillips-curve comparisons (four ML models × three horizons, real-time condition); the full set of thirty-six comparisons, including all twenty-four real-time-versus-revised tests, is reported in Appendix Table A1. Of the twelve ML-versus-Phillips-curve comparisons, only one reaches conventional significance: Gradient Boosting outperforms the Phillips curve at the 6-month horizon (DM = −1.671, p = 0.097), consistent with the RMSFE and out-of-sample $R^2$ patterns noted above. No other ML-versus-Phillips-curve comparison is significant at the 10 percent level in either direction. Among the twenty-four real-time-versus-revised comparisons (testing whether the RSI itself is statistically distinguishable from zero), none is significant at conventional levels once one numerical anomaly, the h = 12 SVR comparison, is set aside as a likely artifact of near-zero-variance loss differentials rather than a genuine finding; we discuss this explicitly in Section 7.

*Table 7. Diebold–Mariano test results: machine learning models vs. the Phillips curve (real-time condition).*

| Comparison | Horizon | DM statistic | p-value | Sig. |
|---|---|---|---|---|
| Random Forest vs. Phillips curve | h = 3 | +1.448 | 0.150 | |
| Gradient Boosting vs. Phillips curve | h = 3 | +0.458 | 0.648 | |
| Elastic Net vs. Phillips curve | h = 3 | +1.608 | 0.110 | |
| SVR vs. Phillips curve | h = 3 | +1.535 | 0.127 | |
| Random Forest vs. Phillips curve | h = 6 | −0.966 | 0.336 | |
| Gradient Boosting vs. Phillips curve | h = 6 | −1.671 | 0.097 | * |
| Elastic Net vs. Phillips curve | h = 6 | +1.356 | 0.177 | |
| SVR vs. Phillips curve | h = 6 | +0.199 | 0.842 | |
| Random Forest vs. Phillips curve | h = 12 | −0.186 | 0.852 | |
| Gradient Boosting vs. Phillips curve | h = 12 | −1.222 | 0.224 | |
| Elastic Net vs. Phillips curve | h = 12 | +0.608 | 0.544 | |
| SVR vs. Phillips curve | h = 12 | −1.306 | 0.194 | |

** $p < 0.10$. No comparison reaches the $p < 0.05$ threshold. Positive DM statistics indicate larger squared errors for the machine learning model (i.e., the Phillips curve performing relatively better); negative statistics indicate the reverse.*

## 5.5 Feature Importance: Real-Time versus Revised, Out-of-Sample SHAP

Table 8 reports out-of-sample mean absolute SHAP values for the Random Forest model at the 3-month horizon, computed using the walk-forward procedure in equation (19). These values are aggregated across per-period explanations of the model that genuinely existed at each recursive forecast origin, not a single in-sample fit. While current CPI inflation and payroll growth dominate under both conditions, their relative weighting shifts substantially, and one feature in particular, PCE inflation, changes role entirely. In the real-time model, PCE inflation ranks sixth among nine features in importance (mean |SHAP| = 0.039), contributing almost nothing to predictions. In the revised-data model, the identical feature ranks first (0.778), an approximately twenty-fold increase in importance, while current CPI inflation's importance falls correspondingly, from 0.834 (rank 1) in the real-time model to 0.223 (rank 2) in the revised model.

This finding is, to the authors' knowledge, novel in the real-time forecasting literature: a machine learning model's reliance on a specific, economically meaningful predictor (the Federal Reserve's preferred inflation gauge) can be substantially a hindsight artifact, invisible to aggregate RMSFE and out-of-sample comparisons, which showed only modest differences between the two data conditions for this model (Table 3, Table 6). A plausible explanation, consistent with the revision statistics in Table 2, is that PCE inflation is one of the more heavily revised series in the sample (standard deviation of revision 3.18pp, nearly as large as GDP's); its first-released values are comparatively uninformative, so the model trained on real-time data learns to discount PCE accordingly and relies instead on the far less-revised CPI signal. The model trained on fully revised data has no such reason to discount PCE and instead exploits its genuine predictive content, which, mechanically, only becomes visible after revision and that a real-time forecaster could never have accessed at the moment of forecasting.

*Table 8. Out-of-sample SHAP feature importance comparison, Random Forest, h = 3.*

| Feature | Mean \|SHAP\| (RT) | Rank (RT) | Mean \|SHAP\| (Rev.) | Rank (Rev.) | Rank shift |
|---|---|---|---|---|---|
| Current CPI inflation ($\pi_t$) | 0.834 | 1 | 0.223 | 2 | −1 |
| Payroll growth (YoY) | 0.332 | 2 | 0.214 | 3 | −1 |
| Unemployment rate | 0.107 | 3 | 0.072 | 5 | −2 |
| 10Y–2Y Treasury spread | 0.075 | 4 | 0.062 | 6 | −2 |
| GDP growth (YoY) | 0.059 | 5 | 0.117 | 4 | +1 |
| PCE inflation (YoY) | 0.039 | 6 | 0.778 | 1 | +5 |
| Inflation, 1 lag | 0.031 | 7 | 0.019 | 9 | −2 |
| Inflation, 2 lags | 0.026 | 8 | 0.029 | 7 | +1 |
| Unemployment gap | 0.019 | 9 | 0.020 | 8 | +1 |

### 5.5.1 Robustness: In-Sample versus Out-of-Sample SHAP

To confirm that the out-of-sample discipline in Section 4.7 is not a cosmetic change, Table 9 reports the same comparison computed the conventional (in-sample) way, a single Random Forest fit on the full real-time or revised panel, explained over that same panel directly alongside the out-of-sample values from Table 9. Two results are notable. First, the qualitative finding survives, and is in fact *stronger* under the correct out-of-sample procedure: PCE inflation's importance swings by a factor of roughly 14 in-sample (0.035 to 0.546) but by a factor of roughly 20 out-of-sample (0.039 to 0.778). This is reassuring evidence that the central finding of this paper is not an artifact of in-sample overfitting, but if anything is attenuated by it. Second, several individual feature importances differ non-trivially between the two procedures, most visibly the unemployment rate, whose real-time importance rises from 0.025 in-sample to 0.107 out-of-sample indicating that in-sample SHAP estimates should not be treated as reliable point estimates of a feature's genuine forecasting relevance even when the qualitative ranking is similar.

*Table 9. In-sample versus out-of-sample SHAP, Random Forest, h = 3.*

| Feature | RT, in-sample | RT, out-of-sample | Rev., in-sample | Rev., out-of-sample |
|---|---|---|---|---|
| Current CPI inflation ($\pi_t$) | 0.761 | 0.834 | 0.331 | 0.223 |
| Payroll growth (YoY) | 0.655 | 0.332 | 0.600 | 0.214 |
| PCE inflation (YoY) | 0.035 | 0.039 | 0.546 | 0.778 |
| GDP growth (YoY) | 0.072 | 0.059 | 0.042 | 0.117 |
| Unemployment rate | 0.025 | 0.107 | 0.012 | 0.072 |
| 10Y–2Y Treasury spread | 0.021 | 0.075 | 0.030 | 0.062 |
| Unemployment gap | 0.019 | 0.019 | 0.010 | 0.020 |
| Inflation, 1 lag | 0.011 | 0.031 | 0.020 | 0.019 |
| Inflation, 2 lags | 0.008 | 0.026 | 0.015 | 0.029 |

## 5.6 Case Study: The 2021–2022 Inflation Peak

Table 10 examines each model's 3-month-ahead forecast for the single month in our evaluation window with the highest realized inflation: March 2022, when year-over-year CPI inflation reached 8.98 percent. Under both data conditions, every model substantially underestimated this peak, a pattern consistent with the well-documented difficulty all forecasting approaches faced during the 2021–2022 inflation surge. The random walk produced the smallest error under both conditions (0.42pp real-time, 0.41pp revised), while SVR produced by far the largest (2.53pp real-time, 2.89pp revised). Notably, the ranking of models by error size is nearly identical across the two data conditions, and the real-time forecasts were, if anything, marginally closer to the eventual outturn for most models, the opposite of what a naive "ML benefits from hindsight" story would predict and consistent with the generally small and insignificant RSI values reported in Section 5.3.

*Table 10. Model forecasts for the 3-month-ahead inflation peak (target month: March 2022, actual = 8.98%).*

| Model | Real-time forecast | RT error | Revised forecast | Rev. error |
|---|---|---|---|---|
| Random walk | 8.56 | 0.42 | 8.57 | 0.41 |
| ADL-OLS | 8.26 | 0.72 | 8.12 | 0.86 |
| AR(1) | 8.25 | 0.73 | 8.25 | 0.73 |
| Phillips curve | 8.24 | 0.74 | 8.25 | 0.73 |
| Gradient Boosting | 7.91 | 1.07 | 8.01 | 0.97 |
| Elastic Net | 7.76 | 1.22 | 7.76 | 1.22 |
| Random Forest | 7.29 | 1.69 | 7.38 | 1.60 |

| Model | Real-time forecast | RT error | Revised forecast | Rev. error |
|---|---|---|---|---|
| SVR | 6.45 | 2.53 | 6.09 | 2.89 |

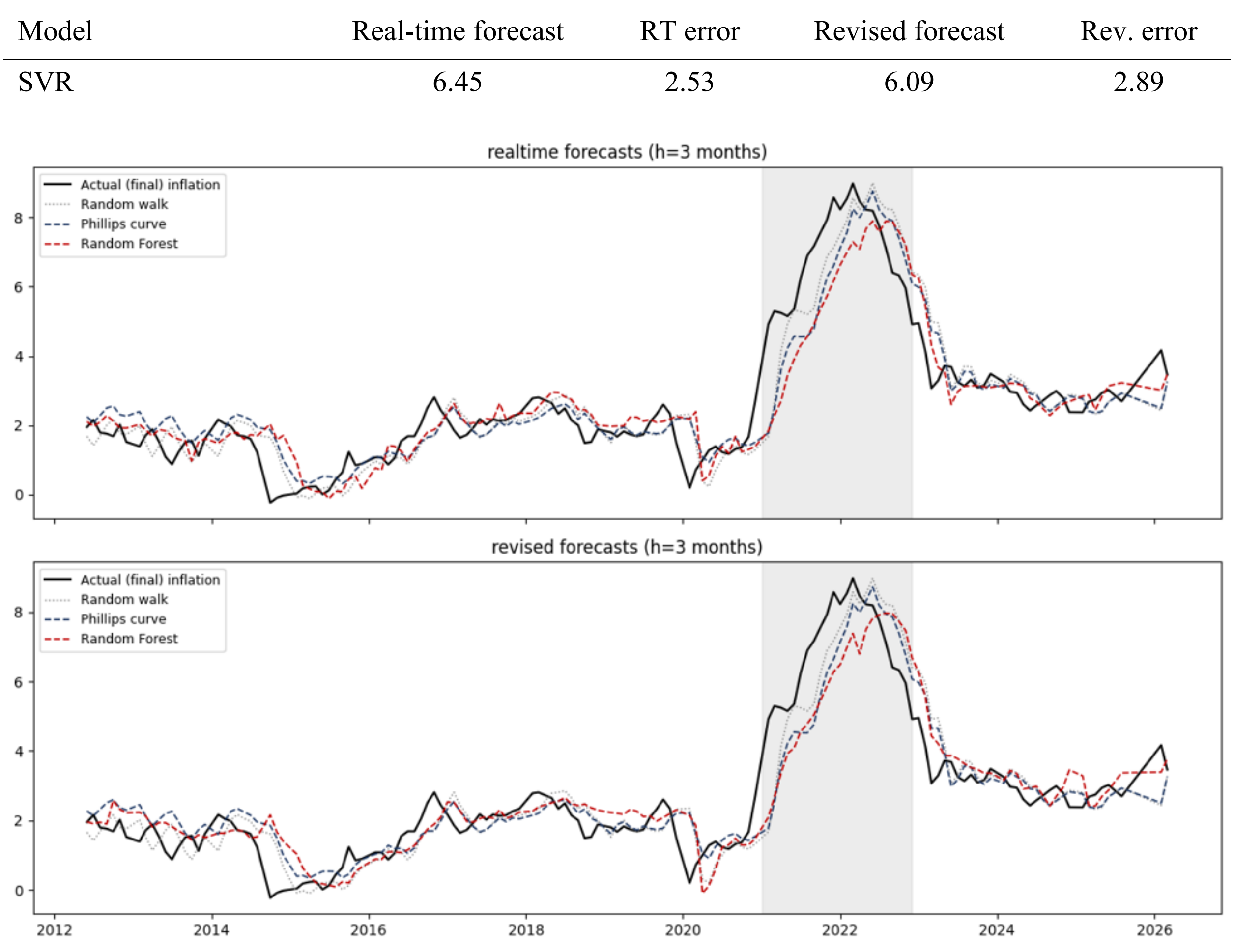

**Figure 5.** *Three-month-ahead forecasts against realized inflation, real-time vs. revised data conditions.*

## 6. Discussion

Three findings merit emphasis. First, at the point-forecast level, the hypothesis that ML's advantage over the Phillips curve is a hindsight artifact is not strongly supported by the evidence presented here: RSI values are generally small, and only one of twelve ML-versus-Phillips-curve comparisons is statistically significant. That exception (Gradient Boosting outperforming under real-time conditions) runs counter to the hindsight-artifact hypothesis, if anything. This is a substantively informative negative result: it suggests that, at least for this feature set, horizon range, and time period, evaluating ML inflation forecasts on revised data, as the majority of the literature does, including Agyekum (2026) and Nortey et al. (2025), does not appear to produce a substantially misleading picture of relative point-forecast accuracy.

Second, and consistent with Atkeson and Ohanian (2001) and with the random-walk puzzle documented by Agyekum et al. (2026) in the context of exchange-rate forecasting, a naive random-walk forecast remains extremely difficult to beat at short and medium horizons, under both data conditions. This consistency across domains the same puzzle appearing in CAD/USD exchange-rate forecasting and in U.S. inflation forecasting suggests it reflects a general property of highly persistent, near-unit-root macro-financial time series rather than an artifact specific to either application. It is conjectured that this property may generalize to other near-unit-root macro-financial series, short-horizon commodity prices, and certain interest-rate spreads among them, though testing this conjecture formally is beyond the scope of the present study.

Third, and most notably, aggregate accuracy metrics mask an economically meaningful divergence in model behavior, and this divergence is, if anything, understated by naive (in-sample) interpretability methods. The out-of-sample SHAP results in Section 5.5 show that a Random Forest trained on revised data effectively substitutes hindsight-informed PCE inflation for the current-inflation signal that the same model, trained on real-time data, relies on almost exclusively. The magnitude of this substitution is larger under the methodologically correct walk-forward procedure (a roughly twenty-fold swing) than under the naive in-sample alternative (roughly fifteen-fold; Table 9). Because PCE and CPI inflation are highly correlated at low frequencies, this substitution has only a modest effect on point-forecast accuracy, but it means that a practitioner inspecting a revised-data-trained model's feature importances to understand what drives the forecast would draw a materially incorrect conclusion about which variables a real-time system could actually exploit. This has a direct practical implication: interpretability analyses of ML macroeconomic forecasting models should be conducted out-of-sample, using the same walk-forward discipline applied to accuracy evaluation, and on real-time-consistent training data whenever the resulting narrative is intended to inform real-time forecasting practice or policy communication.

These three findings, taken together, suggest a nuanced answer to the paper's central question. Machine learning's forecasting advantage over the Phillips curve, where it exists at all in the results presented here (essentially confined to Gradient Boosting at 6- and 12-month horizons), survives the transition to real-time data largely intact. What does not survive intact is the interpretive account of why the model performs as it does: the identity of the predictors driving the forecast shifts materially between data conditions, even when the resulting forecast accuracy does not.

# 7. Limitations and Robustness

Several limitations qualify our findings. First, our sample spans 2000–2026 and yields at most 208 monthly out-of-sample forecasts at the shortest horizon, with fewer at longer horizons. This is a modest sample for detecting statistically significant differences via the Diebold–Mariano test, and our null results on the RSI should be interpreted as "not detected in this sample" rather than "conclusively absent." Statistical power to detect a given true RSI is correspondingly limited, particularly at the 12-month horizon, where only 204 observations are available and forecast errors exhibit substantial serial correlation from the overlapping-horizon construction.

Second, we identified one numerical anomaly in our full results output, reported in Appendix Table A1: the real-time-versus-revised Diebold–Mariano statistic for SVR at the 12-month horizon returned an implausibly large value (DM ≈ 394,000, nominal $p < 0.001$), almost certainly reflecting a near-zero-variance loss-differential series. In other words, real-time and revised SVR forecast errors were numerically almost identical for most of the sample, causing the variance term in the denominator of equation (16) to collapse toward the numerical floor imposed in estimation rather than yielding a genuine, economically meaningful result. We flag this comparison as not significant and recommend that any subsequent revision of this analysis re-estimate it using a bootstrap-based variant of the Diebold–Mariano test, which is more robust to this variance-degeneracy failure mode than the asymptotic Newey–West-type correction used here.

Third, our target variable is defined using the fully revised ("final") value of future inflation in both the real-time and revised experiments; an alternative design would define the real-time experiment's target using the first-released value instead, which would more strictly separate "what could have been known" from "what we now know to be true" and would serve as a natural robustness check for future work, particularly given that CPI's own revisions are small (Table 2) and unlikely to materially alter target values regardless of which convention is used.

Fourth, we consider a single ML architecture family per model class at default-adjacent hyperparameters; more extensive hyperparameter tuning, particularly for SVR (whose RMSFE was uniformly the weakest of the eight models across all horizons and conditions), could alter the relative ranking of ML models without necessarily changing the paper's central real-time-versus-revised comparison. Relatedly, our recursive re-estimation procedure holds hyperparameters fixed across the expanding window rather than re-tuning them at each forecast origin; a fully time-varying hyperparameter search, while considerably more computationally expensive, would be a natural extension.

Fifth, the out-of-sample SHAP procedure in Section 4.7 is markedly more computationally expensive than the in-sample alternative, since it requires re-fitting and re-explaining a separate model at every one of roughly 150 recursive steps rather than once; this cost currently confines the analysis to a single model (Random Forest) at a single horizon (h = 3). Extending the out-of-sample interpretability comparison to Gradient Boosting, the model with the clearest real-time-versus-revised accuracy divergence in Table 4 and Table 5, and the paper's only model with consistent positive out-of-sample $R^2$ at longer horizons (Table 6) and to the 6- and 12-month horizons is a natural next step, and one we would expect, based on the RSI and out-of-sample $R^2$ patterns already documented, to reveal an interpretability picture at least as horizon-dependent as the accuracy results themselves.

# 8. Conclusion

This paper examined whether machine learning's forecasting advantage over the traditional Phillips curve survives when both are evaluated on real-time, vintage-consistent data rather than the fully revised data used in almost all published comparisons, including Agyekum (2026), and whether ML feature-importance rankings survive an analogous transition from in-sample to genuinely out-of-sample estimation. For point-forecast accuracy, the answer is broadly reassuring: differences between real-time and revised evaluation are generally small and, with one exception, statistically insignificant across thirty-six formally tested model–horizon comparisons; out-of-sample cap R squared confirms that only Gradient Boosting delivers consistent skill beyond the random-walk benchmark, and that this skill is concentrated at longer horizons; and a naive random walk remains a difficult benchmark to beat, regardless of data vintage, consistent with the findings of Agyekum et al. (2026) on exchange-rate forecasting. The more consequential finding, however, concerns interpretation rather than accuracy: a fully walk-forward, out-of-sample SHAP analysis reveals that models trained on revised data rely heavily on predictors, in this case PCE inflation, whose apparent importance is substantially a hindsight artifact rather than a real-time signal. This finding is, if anything, strengthened rather than weakened once the naive in-sample SHAP procedure common in the literature is replaced with genuine walk-forward estimation. As machine learning tools become more embedded in central bank and market forecasting practice, it is argued that accuracy, significance, and interpretability should all be audited under real-time-consistent, out-of-sample conditions: the first two appear, in this application, not to materially alter the assessment relative to the revised-data convention; the third can alter it substantially. The formal Revision Sensitivity Index, the out-of-sample SHAP procedure, and the vintage-consistent evaluation pipeline developed here are proposed as reusable tools for future work extending this analysis to additional model classes, forecast targets, and economies.

# Appendix A. Full Diebold–Mariano Test Results

Table A1 reports all thirty-six Diebold–Mariano comparisons underlying Section 5.4: twelve machine-learning-versus-Phillips-curve comparisons (real-time condition) and twenty-four real-time-versus-revised comparisons (one per model per horizon).

*Table A1a. Machine learning vs. Phillips curve, real-time condition (all horizons).*

| Model | Horizon | DM statistic | p-value |
|---|---|---|---|
| Random Forest | h = 3 | +1.448 | 0.150 |

| Model | Horizon | DM statistic | p-value |
|---|---|---|---|
| Gradient Boosting | h = 3 | +0.458 | 0.648 |
| Elastic Net | h = 3 | +1.608 | 0.110 |
| SVR | h = 3 | +1.535 | 0.127 |
| Random Forest | h = 6 | −0.966 | 0.336 |
| Gradient Boosting | h = 6 | −1.671 | 0.097* |
| Elastic Net | h = 6 | +1.356 | 0.177 |
| SVR | h = 6 | +0.199 | 0.842 |
| Random Forest | h = 12 | −0.186 | 0.852 |
| Gradient Boosting | h = 12 | −1.222 | 0.224 |
| Elastic Net | h = 12 | +0.608 | 0.544 |
| SVR | h = 12 | −1.306 | 0.194 |

*Table A1b. Real-time vs. revised, same model (all models, all horizons).*

| Model | Horizon | DM statistic | p-value |
|---|---|---|---|
| Random walk | h = 3 | +1.199 | 0.233 |
| AR(1) | h = 3 | +0.841 | 0.402 |
| Phillips curve | h = 3 | +0.121 | 0.904 |
| ADL-OLS | h = 3 | +0.858 | 0.392 |
| Random Forest | h = 3 | +0.582 | 0.561 |
| Gradient Boosting | h = 3 | +0.527 | 0.599 |
| Elastic Net | h = 3 | +1.411 | 0.161 |
| SVR | h = 3 | +1.069 | 0.287 |
| Random walk | h = 6 | +0.815 | 0.417 |
| AR(1) | h = 6 | −0.178 | 0.859 |
| Phillips curve | h = 6 | −1.324 | 0.188 |
| ADL-OLS | h = 6 | +1.208 | 0.229 |
| Random Forest | h = 6 | −0.560 | 0.576 |
| Gradient Boosting | h = 6 | −0.772 | 0.441 |
| Elastic Net | h = 6 | +0.969 | 0.334 |
| SVR | h = 6 | +0.834 | 0.406 |
| Random walk | h = 12 | −0.995 | 0.322 |

| Model | Horizon | DM statistic | p-value |
|---|---|---|---|
| AR(1) | h = 12 | −0.160 | 0.873 |
| Phillips curve | h = 12 | −0.983 | 0.328 |
| ADL-OLS | h = 12 | +1.001 | 0.319 |
| Random Forest | h = 12 | +1.186 | 0.238 |
| Gradient Boosting | h = 12 | −1.214 | 0.227 |
| Elastic Net | h = 12 | +0.997 | 0.321 |
| SVR | h = 12 | +394333.5† | <0.001† |

** $p < 0.10$. † Flagged as a numerical anomaly attributable to near-zero-variance loss differentials; see Section 7 for discussion. This result should not be interpreted as a genuine finding without further robustness testing.*

# Appendix B. Model Hyperparameters

*Table B1. Hyperparameter settings for all estimated models.*

| Model | Key hyperparameters |
|---|---|
| Random walk | None (no estimation) |
| AR(1) | OLS, no regularization |
| Phillips curve | OLS, no regularization |
| ADL-OLS | OLS, no regularization, 9 features |
| Random Forest | 200 trees; max depth 4; min. leaf size 5 |
| Gradient Boosting | 150 stages; max depth 2; learning rate 0.05 |
| Elastic Net | $\alpha = 0.5$; $\lambda = 0.1$; standardized features |
| SVR | RBF kernel; $C = 1.0$; $\varepsilon = 0.1$; default bandwidth |